\pdfoutput=1
\documentclass[prd,aps,preprint,amsmath,nofootinbib,amssymb,eqsecnum,showkeys,tightenlines]{revtex4}
\usepackage{amsmath, latexsym, amssymb, graphicx, color, multirow}
\usepackage{graphicx, bm}
\usepackage{subfigure}
\usepackage{mathrsfs}
\usepackage{verbatim}
\usepackage[final]{pdfpages}
\usepackage[sc]{mathpazo}
\usepackage{graphicx}
\usepackage{color}
\usepackage{epstopdf}
\usepackage{subfigure}
\usepackage{hyperref}

\newcommand{\pythia}{\textsc{Pythia8}~}
\newcommand{\delphes}{\textsc{Delphes-3.4.0}~}
\newcommand{\fastjet}{\textsc{FastJet-3.2.1}~}
\newcommand{\nnpdf}{\textsc{NNPDF23LO}~}

\newcommand{\madgraph}{\textsc{MG5\_}a\textsc{MC@NLO v2.6.0}~}

\begin{document}
\title{Least constrained supersymmetry with R-parity violation}

\author{Jinmian Li$^{a}$}
\email[]{jmli@scu.edu.cn}

\author{Tianjun Li$^{b,c}$}
\email[]{tli@itp.ac.cn}

\author{Wenxing Zhang$^{b,c}$}
\email[]{zhangwenxing@itp.ac.cn}

\affiliation{$^a$ College of Physical Science and Technology, Sichuan University, Chengdu, Sichuan 610065, China}
\affiliation{$^b$ CAS Key Laboratory of Theoretical Physics, Institute of Theoretical Physics, Chinese Academy of Sciences, Beijing 100190, China}
\affiliation{$^c$ School of Physical Sciences, University of Chinese Academy of Sciences,
No.~19A Yuquan Road, Beijing 100049, China}

\begin{abstract}
The strong constraints on the R-parity conserving supersymmetry (SUSY) from the LHC searches motivate us to consider 
the new models in which the low-scale SUSY is still allowed. We propose a kind of R-parity violating SUSY 
scenario with a nonzero $U^c_2 D^c_2 D^c_3$ operator. 
Three relevant LHC searches are recast to test the status of this scenario in terms of four simplified models, with either light stop-Bino, stop-Higgsino, sbottom-Bino, or sbottom-Higgsino. 
Some difficult scenarios for the LHC SUSY searches in these simplified models are identified. 
By extrapolating the current LHC searches to the future 14 TeV LHC with integrated luminosity of 3000 fb$^{-1}$, 
the stop/sbottom masses in all scenarios can be probed up to $\sim 800$-$1100$ GeV. 

\end{abstract}

\maketitle

\section{Introduction}\label{sec:intro}

As one of the most promising candidates for new physics beyond the Standard Model (SM), 
supersymmetry (SUSY)~\cite{Nilles:1983ge, Haber:1984rc} provides an elegant solution to the gauge hierarchy problem.
In the supersymmetric SMs (SSMs), the gauge coupling unification can be realized. 
In order to forbid the renormalizable superpotential terms that violate the baryon number ($B$) and lepton number ($L$) and thus, induce the fast proton decays,
the $Z_2$ $R$-parity ($R=(-1)^{(3B-L)+2S}$) is introduced, where $S$ is
the particle spin~\footnote{There will be a higher dimensional $B$ and $L$ violating superpotential term $\mathcal{W} \subset U^c_i D^c_j D^c_k  E^c_l/ \Lambda$~\cite{Sakai:1981pk,Weinberg:1981wj}, which respects the $R$-parity. Assuming a large cutoff ($\Lambda$) at around GUT scale, one can satisfy the experimental bounds on the proton decay lifetime.}.
Under the $R$-parity symmetry, all the SM particles are even while their superpartners are odd. 
Thus, the lightest supersymmetric particle (LSP) will be stable. Especially, the neutralino LSP serves 
as the very promising weakly-interacting-massive-particle dark matter (DM) candidate, which can have
the correct DM relic density as well~\cite{Jungman_1995df}. 

However, the searches for $R$-parity conserving (RPC) SUSY signals at the LHC, which mainly rely on 
the large missing transverse energy (MET) in the final state, have given quite strong constraints. 
The gluino/squark masses have been pushed to a couple of TeV~\cite{Aaboud:2017vwy,Sirunyan:2018vjp}, 
challenging the naturalness problem~\cite{Hall_2011aa,Papucci:2011wy} and little hierachy problem~\cite{BasteroGil:2000bw,Bazzocchi:2012de} of the SUSY theories. On the other hand, the main goal for SUSY is to solve the gauge hierarchy problem,
so $R$-parity is not mandatory. 
The renormalizable $R$-parity violation (RPV) terms in the superpotential are~\cite{Hall_1983id}
\begin{equation}\label{wrpv}
\mathcal{W}= \mu'_i L_i H_u + \lambda_{ijk} L_i L_j E^c_k + \lambda_{ijk}'L_i Q_j D^c_k + \lambda_{ijk}'' U^c_i D^c_j D^c_k  
\end{equation}
where $L_i$, $E^c_i$, $Q_i$, $U^c_i$, $D^c_i$, and $H_u$ denote the left-handed lepton, right-handed lepton, left-handed doublet quarks, right-handed up-type quarks, right-handed down-type quarks, and up-type Higgs. The $\lambda$ and $\lambda$'' are antisymmetric in the exchange of $i\to j$ and $j \to k$, respectively. In Eq.~(\ref{wrpv}), the first three terms break the lepton number symmetry and the last term breaks the baryon number symmetry. Note that the proton can still be stable 
as long as only the lepton number or baryon number symmetry is broken. 

The RPV SUSY has been searched at the LHC in several different channels (for recent reviews,
see Refs.~\cite{Franceschini:2015pka,Redelbach_2015meu}), with special attention paid to the gluino and top squark productions. The signature of a pure hadronic multijet~\cite{Aaboud_2018lpl} in the final state has been searched to constrain the gluino pair production if $\lambda''_{ijk, ~i \neq 3}$ is nonzero. When $\lambda_{3jk}'' \neq 0$, there could be top quarks from the gluino decay, the leptonic decay of which gives leptons + multijet final state~\cite{Aaboud_2017faq,Sirunyan_2017dhe,Aaboud_2017dmy}. Searches for the same final state are also constraining the $L_i Q_j D^c_k$ operator. These operators will also lead to stop either decaying into two jets or decaying into a lepton and a jet, which has been searched in resonant dijet pair~\cite{Aaboud_2017nmi} and lepton-jet pair~\cite{Aaboud_2017opj}. 
Finally, if the RPV couplings are small such that the $R$-hadrons are stable at the scale of the detector size, there are searches for long-lived $R$-hadrons~\cite{ATLAS_2018yey}. 
From those searches, we can observe that the bounds obtained for those operators giving leptons in the final state are quite stringent: e.g., a gluino being excluded up to $\sim 2$ TeV, stops being excluded up to $\sim 1$ TeV. We note that such bounds may be relaxed to some extent by extending the decay chain with extra particles~\cite{Evans:2013jna,Asano:2014aka}, due to the soft final states. As a result, these scenarios will be also in tension with the naturalness problem, same as for the RPC case. But the bounds with $U^c_i D^c_j D^c_k$ operator are much weaker due to the heavy hadronic activity expected at the LHC~\cite{Allanach:2012vj,Durieux:2013uqa,Bhattacherjee:2013gr,Graham:2014vya,Diglio:2016ynj,Buckley:2016kvr,Evans:2018scg}.  There are plenty of studies that try to improve the sensitivity for searching the RPV scenario with a nonzero $\lambda''_{ijk}$ , by using jet substructure analysis on either neutralino jet~\cite{Butterworth_2009qa} or top squark jet~\cite{Bai_2013xla,Bhattacherjee_2013tha}, and by multivariate analyses~\cite{Bardhan_2016gui}.  

Among all possible $\lambda''_{ijk}$ , the scenario with $i=3$ will give a top quark in the final state from the on-shell/off-shell neutralino decay. The leptonic mode of which will be stringently constrained. Moreover, terms with $i,j,k = 1,2$ are constrained~\cite{Barbier_2004ez} by the low energy experiments such as single nucleon decay channels and neutron-antineutron oscillation. In this paper, we will consider the least constrained scenario, i.e., RPV dominated by a nonzero $\lambda''_{223}$~\footnote{For collider phenomenology, those subdominant couplings are not relevant as long as they are not contributing much to the production processes and sparticles decays. The single coupling dominance ansatz has been adopted in many other similar studies~\cite{Evans:2012bf,Franceschini:2012za,Berger:2013sir,Duggan:2013yna,Aaboud_2017faq}.}. 
Considering the renormalization group equation of $Y = (\lambda^{\prime \prime 2}_{212} +\lambda^{\prime \prime 2}_{213} + \lambda^{\prime \prime 2}_{223})/ 4 \pi$, the requirement of perturbativity of $Y$ at the unification scale (i.e., $Y<1$) gives the only constraint on $\lambda''_{223}$,  {\it i.e.,} $\lambda''_{223} < 1.25$ at the electroweak scale~\cite{Goity:1994dq}. 
In some experimental searches as well as phenomenological studies, the bounds on the top/bottom squark with RPV were studied under the assumption that they are the LSP and 100\% decay through the RPV operator~\cite{Choudhury:2011ve,Brust:2012uf,Han:2012cu,Franceschini:2012za,Berger:2013sir,Aaboud_2017dmy,Aaboud:2017nmi}. However, this is not valid in the traditional
supersymmetry breaking scenarios; for example,  the SSMs inspired by a grant unified theory (GUT) 
with gravity mediation~\cite{Kowalska_2015zja,Dercks:2017lfq}, where the lightest neutralino could be the LSP, etc.
Reference~\cite{Evans:2012bf} performed a systematic study of LHC run-I coverage of all trilinear RPV operators in the pair production of light stops. 
In particular, the bound on the stop pair production with the subsequent decay through intermediate Bino or Higgsino, which further decays into jets by a $U^c_2 D^c_1 D^c_3$ operator, are considered. And a similar process with Bino/Higgsino decays through the $U^c_3 D^c_2 D^c_3$ operator is searched by the ATLAS Collaboration~\cite{Aaboud_2017faq}.
The sensitivity of an upgraded LHC on those RPV operators were studied in Ref.~\cite{Duggan:2013yna}, which includes the case of $\tilde{t} \to t \tilde{B} \to t (jjj)$ through the $U^c_2 D^c_1 D^c_2$ operator. 
We will study the top/bottom squark bounds in the cases where either there is a Bino or Higgsino LSP. Then, the top squark can only decay into on-shell/off-shell top quark and a neutralino, which further decay through 
the RPV operator $U^c_2 D^c_2 D^c_3$, {\it i.e.}, $\tilde{\chi}^0 \to csb$~\footnote{This case is similar to that in Refs.~\cite{Evans:2012bf,Duggan:2013yna}. But we will perform our analysis on the two-dimensional $m_{\tilde{t}} - m_{\tilde{\chi}^0}$ parameter plane.}. While the bottom squark decay is more complicated, besides the RPC decay of $\tilde{b} \to b \tilde{\chi}^0$/$\tilde{b} \to t^* \tilde{\chi}^\pm$, there is also a direct RPV decay $\tilde{b} \to cs$. The LHC bounds on the mixture of these branching ratios will be considered in this work.

This paper is organized as follows. In Sec.~\ref{sec:model}, we introduce four simplified SUSY models with a $U^c_2 D^c_2 D^c_3$ R-parity violating operator. Their corresponding LHC signals will be discussed. In Sec.~\ref{sec:lhc}, the current LHC sensitivities to those signals as well as their future prospects are studied. Our conclusions are given in Sec.~\ref{sec:conclude}. We also show the validation of our recasting of experimental searches in Appendixes~\ref{app1}-~\ref{app3}.

\section{The simplified models and signals} \label{sec:model}
We consider the simplified RPV SUSY models with following assumptions:
\begin{itemize}
\item $\lambda''_{223}$ is the only nonvanishing RPV coupling. 
\item The only light colored particle is a mostly right-handed bottom squark or top squark, while 
all the others are too heavy to be produced at the LHC. 
\item  Inspired from SUSY GUT as well as SUSY naturalness, we assume there is either a bino ($\tilde{B}$) or a Higgsino ($ \tilde{H}$) that has a mass below the sbottom/stop, acting as the LSP. 
\end{itemize}
As a result, we have four versions of simplified models: $\tilde{t} - \tilde{B}$, $\tilde{t} - \tilde{H}$, $\tilde{b}-\tilde{B}$, $\tilde{b} - \tilde{H}$. 

In the minimal SUSY framework, the tree-level mass matrix of the neutralino sector in the basis of $(\tilde{B}, \tilde{W^0}, \tilde{H_d^0}, \tilde{H_u^0})$ is 
\begin{equation}\label{eq:Y} \small
\mathcal{M} =\left( \begin{array}{cccc}
M_1 & 0 & -m_Z \cos\beta\sin\theta_W & m_Z\sin\beta\sin\theta_W \\
0   & M_2 & m_Z\cos\beta\cos\theta_W & -m_Z\sin\beta\cos\theta_W \\
-m_Z\cos\beta\sin\theta_W & m_Z\cos\beta\cos\theta_W & 0 & -\mu \\
m_Z\sin\beta\sin\theta_W & -m_Z\sin\beta\cos\theta_W & -\mu & 0 \end{array} \right), 
\end{equation}
where $M_1$ and $M_2$ are soft mass parameters for bino and wino, 
$\mu$ is the bilinear Higgs mass in the superpotential, $\tan\beta$ is the ratio 
between the vacuum expectation values of $H_u$ and $H_d$, and $\theta_W$ is the weak mixing angle. 
The limit $M_1 \ll M_2,\mu$ gives the bino LSP in our simplified model, while the Higgsino LSP is more involved. In the limit $\mu \ll M_1,M_2$, there will be two mass eigenstates 
for neutralinos
that have masses close to $\mu$. Both have a similar amount of the $H_u$ and $H_d$ component. Their mass difference at the tree level is given by 
\begin{equation}
M_{\tilde{\chi}_2^0}-M_{\tilde{\chi}_1^0}=\frac{m_Z^2}{2}\left(\frac{\sin^2 \theta_W}{M_1}+\frac{\cos^2 \theta_W}{M_2}\right),
\end{equation}
which is tiny in the decoupling limit $\mu \ll M_1,M_2$. 
Moreover, the Higgsino has another component in the chargino sector. The chargino mass matrix can be written as
\begin{equation}\label{eq:X}
\mathcal{X}=
\left( \begin{array}{cc}
M_2 & \sqrt{2} m_Z\cos\theta_W\sin\beta  \\
\sqrt{2} m_Z\cos\theta_W\cos\beta  & \mu
\end{array} \right) . 
\end{equation}
The mass difference between the charged Higgsino and the lighter neutral Higgsino at the tree level is thus given by 
\begin{equation}
M_{\tilde{\chi}_1^\pm}-M_{\tilde{\chi}_1^0}=\frac{m_Z^2}{2}\left[\sin2\beta \left( \frac{\sin^2\theta_W}{M_1}-\frac{\cos^2\theta_W}{M_2}\right) + \left(\frac{\sin^2\theta_W}{M_1} + \frac{\cos^2\theta_W}{M_2}\right) \right] .
\end{equation}
It can be simplified further in the large $\tan\beta$ limit,
\begin{equation}\label{eq:a}
M_{\tilde{\chi}_1^\pm}-M_{\tilde{\chi}_1^0}=\frac{m_Z^2}{2}\left(\frac{\sin^2 \theta_W}{M_1}+\frac{\cos^2 \theta_W}{M_2}\right) .
\end{equation}
To conclude, we will have two neutralinos and one chargino for the Higgsino LSP cases. All of those three particles have masses close to $\mu$.  According to Eq.~(\ref{eq:a}), the mass difference between heavier Higgsinos and the LSP is less than $\sim \mathcal{O}(1)$ GeV at tree level when the gaugino masses are set to be $M_1, ~M_2 \gtrsim 5$ TeV~\footnote{The electroweak loop correction will induce the mass splitting of $\sim 200$ MeV between charged and neutral Higgsino~\cite{Thomas:1998wy}.}.  
For specification and simplicity, we will take $m_{\tilde{\chi}^\pm_1} = m_{\tilde{\chi}^0_2}= m_{\tilde{\chi}^0_1} +1$ GeV throughout this work. Note that the changing of the mass difference within a few GeV will not affect our results, as long as the soft leptons/jets from the heavier state decays ($\tilde{\chi}^\pm_1(\tilde{\chi}^0_2) \to f \bar{f} \tilde{\chi}^0_1, ~f=\ell,\nu,q$) are undetectable.

\begin{figure}[htb]
\begin{center}
\includegraphics[width=0.5\textwidth]{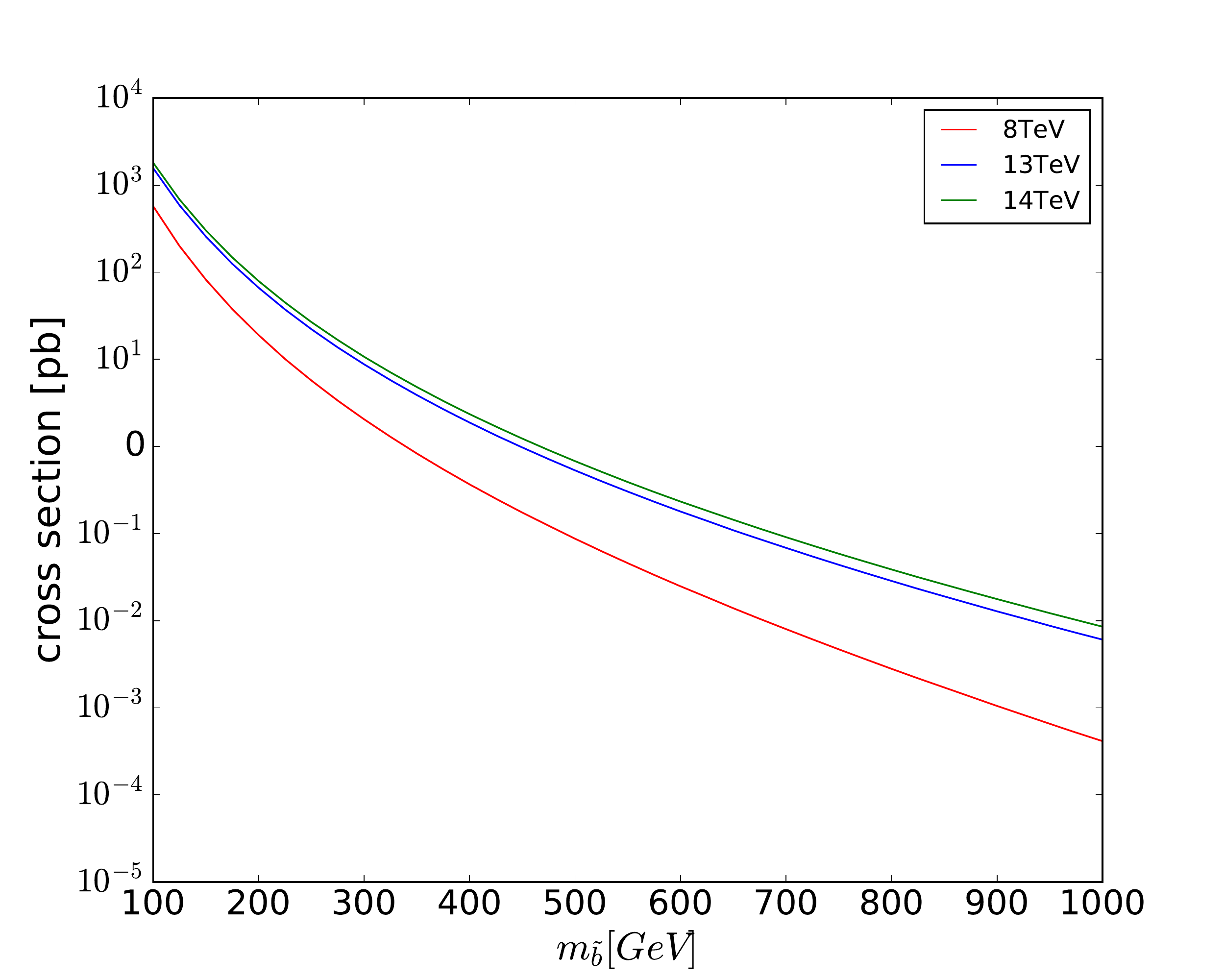}
\end{center}
\caption{\label{fig:xsecs} Bottom squark production cross section at 8 TeV, 13 TeV and 14 TeV proton-proton collider.}
\end{figure}

The dominant SUSY signals of these simplified models at the LHC are the sbottom/stop pair productions with their subsequent decays. Their productions are simply through the QCD couplings, thus with approximatively identical cross section for stop and sbottom. In Fig.~\ref{fig:xsecs}, we plot the next-to-leading order cross sections of sbottom pair production at 8 TeV, 13 TeV and 14 TeV proton-proton collider, which are calculated by Prospino2~\cite{Beenakker:1996ed}. 

\begin{figure}[thb]
\begin{center}
\includegraphics[width=0.3\textwidth]{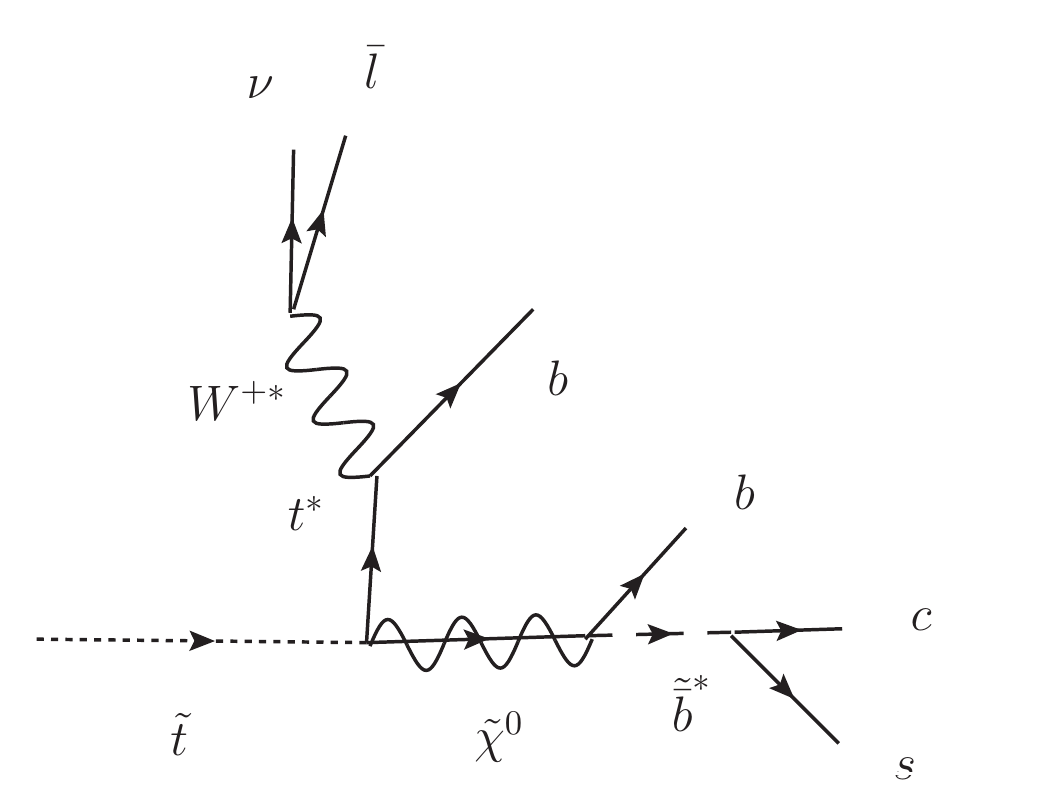}
\includegraphics[width=0.5\textwidth]{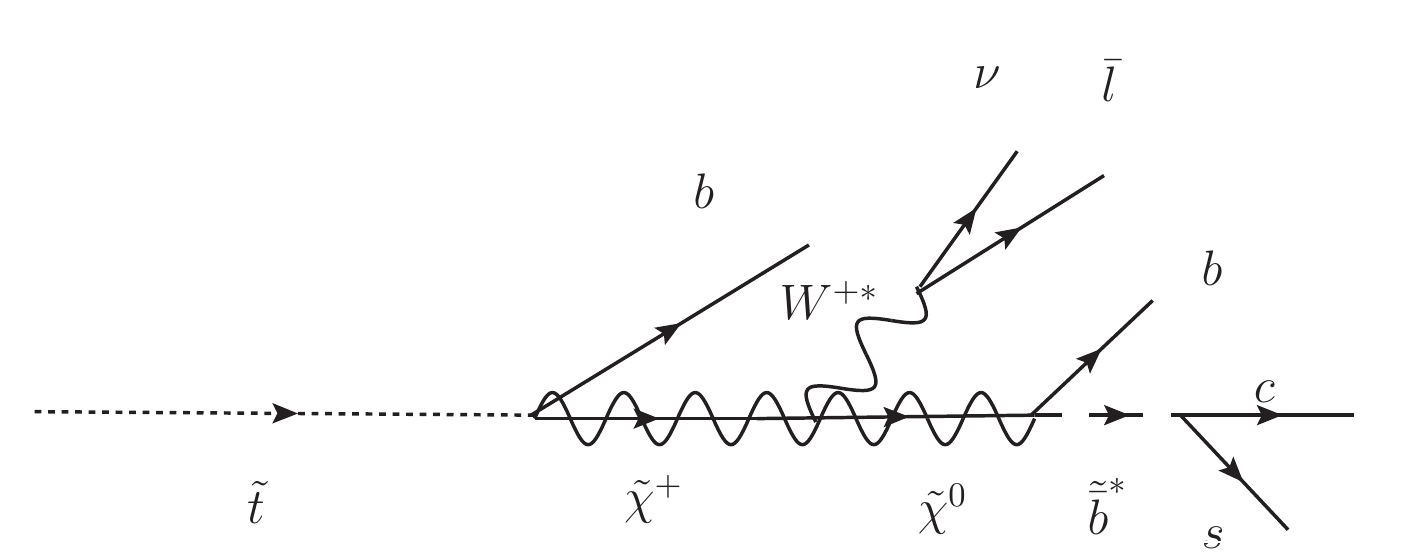}
\end{center}
\caption{\label{fig:stops} Top squark decays channels in our simplified models. }
\end{figure}

The decays of stop/sbottom are more complicated. We will discuss each of the simplified models case by case.
\begin{itemize}
\item $\tilde{t} - \tilde{B}$: The only allowed channel for the stop decay is $\tilde{t} \to t^{(*)} \tilde{\chi}^0$, with $\tilde{\chi}^0=\tilde{B}$ and top quark being either on-shell or off-shell depending on the mass difference between $\tilde{t}$ and $\tilde{\chi}^0$, as shown in the left panel of Fig.~\ref{fig:stops}. 
\item $\tilde{t} - \tilde{H}$: Since there is also a charged Higgsino lighter than the stop, besides the channel $\tilde{t} \to t^{(*)} \tilde{\chi}^0_{1,2}$ , there is a decay of $\tilde{t} \to b \tilde{\chi}^\pm$ with a subsequent decay $\tilde{\chi}^\pm \to W^* \tilde{\chi}^0_1$ shown in the right panel of Fig.~\ref{fig:stops}. When the stop is right-handed dominating, the decay width of each channel is given by
\begin{align}
 \Gamma(\tilde{t}_R \to t \tilde{H}^0_{1,2}) & = \frac{1}{16 \pi m^2_{\tilde{t}}} (\frac{Y_t }{\sqrt{2}})^2 (m^2_{\tilde{t}} - m^2_t - m^2_{\tilde{H}^0_{1,2} } ) \lambda^{1/2}(m^2_{\tilde{t}} , m^2_t, m^2_{\tilde{H}^0_{1,2} }  ) , \label{tth}\\ 
  \Gamma(\tilde{t}_R \to b \tilde{H}^\pm) & = \frac{1}{16 \pi m^2_{\tilde{t}}} (Y_t)^2 (m^2_{\tilde{t}} - m^2_b - m^2_{\tilde{H}^\pm } ) \lambda^{1/2}(m^2_{\tilde{t}} , m^2_b, m^2_{\tilde{H}^\pm }  ),  \label{tbh}
\end{align}
with the two-body phase space function $\lambda(x,y,z) = x^2 + y^2 + z^2 - 2(xy+xz+yz)$ and $Y_t$ is the top quark Yukawa coupling. 
These two channels are comparable if they are kinematically allowed, while the later one is dominating when the mass difference between the stop and Higgsino is small ($m_{\tilde{t}} < m_t + m_{\tilde{H}}$). 

\item $\tilde{b} - \tilde{B}$: Firstly, the sbottom can decay through the RPC channel with decay width
\begin{align}
\Gamma(\tilde{b}_R \to b \tilde{B}^0 ) &=    \frac{1}{16 \pi m^2_{\tilde{b}}} ( \frac{\sqrt{2} e}{3 \cos\theta_W} )^2 (m^2_{\tilde{b}} - m^2_b - m^2_{\tilde{B} } ) \lambda^{1/2}(m^2_{\tilde{b}} , m^2_b, m^2_{\tilde{B} }  ) , \label{bbB} 
\end{align}
where $Y_b$ is the bottom Yukawa coupling. In the mass limit $m_{\tilde{b}} \gg m_b$ and $m_{\tilde{B}^0}$, the decay width can be estimated as $\Gamma(\tilde{b} \to b \tilde{B}^0 ) \sim 0.013 \times \frac{m_{\tilde{b}}}{8 \pi}$.  In contrast to the $\tilde{t} - \tilde{B}$ simplified model, the sbottom can also decay directly through the RPV operator $U^c_2 D^c_2 D^c_3$. Its decay width can be written as
\begin{align}
 \Gamma(\tilde{b} \to \bar{s} \bar{c}) &= \frac{m_{\tilde{b}} }{8 \pi} |\lambda''_{223}|^2.  \label{bsc}
\end{align}
Thus, the decay width in Eqs.~(\ref{bbB}) and (\ref{bsc}) will be around the same size if the $\lambda''_{223} \sim \mathcal{O}(0.1)$. 
\item $\tilde{b} - \tilde{H}$: This case is similar with the $\tilde{t} - \tilde{H}$ simplified model. The sbottom can decay either through $\tilde{b} \to b \tilde{H}^0_{1,2}$ or $\tilde{b} \to t \tilde{H}^\pm$. 
Comparing to Eqs.~(\ref{tth}) and (\ref{tbh}), the decay widths of both channels are proportional to the bottom quark Yukawa coupling instead of the top quark Yukawa coupling, for the pure right-handed sbottom~\footnote{We note that the decay width of left-hand sbottom $\Gamma(\tilde{b}_L \to t \tilde{H}^\pm) \propto Y_t^2$. Because $Y_t \gg Y_b$, even a small component of a left-handed sbottom will lead to $\Gamma(\tilde{b} \to t \tilde{H}^\pm) \gg \Gamma(\tilde{b} \to b \tilde{H}^0_{1,2})$, giving more top quarks in the final state. Considering this, we will give the sbottom a little mixing of the left-handed part and focus on the  $\Gamma(\tilde{b} \to t \tilde{H}^\pm)$ case.  Besides, there is a direct RPV channel $\tilde{b} \to s c$, with its decay width given in Eq.~(\ref{bsc}) as well. }.
\end{itemize} 

In our setup, the neutralino will decay into three-body final states through an off-shell squark ($\tilde{b}/\tilde{s}/\tilde{c}$). We should require that the decay length be within the detector. Otherwise, the neutralino will leave nothing inside the detector, behaving exactly the same as in the RPC case. The RPV three body decay width of a neutralino is~\cite{Monteux:2016gag}

\begin{align}
\Gamma(\tilde{\chi}^0_{1} \to b c s) &= \frac{m^5_{\tilde{\chi}^0_{1}}}{1024 \pi^3 m^4_{\tilde{q}} } |\lambda''_{223}|^2 C^2 \cdot I(m_{\tilde{q}}, m_{\tilde{\chi}^0_1} ), 
\end{align}
where we have assumed that all the quark masses are negligible, the phase space integral 
\begin{align}
 I(m_{\tilde{q}}, m_{\tilde{\chi}^0_1} ) = \int^1_0 \frac{12z^2 (1-z)}{(1- (1-z)\frac{m_{\tilde{\chi}^0_1}^2}{m^2_{\tilde{q}}})^2}
\end{align}
and $C$ is the coupling between the $\tilde{\chi}^0_1- q -\tilde{q}$. For $m_{\tilde{\chi}^0_1} \sim 100$ GeV, $C \sim 0.1$, and  $m_{\tilde{q}} \gg m_{\tilde{\chi}^0_1}$, $|\lambda''_{223}|/m^2_{\tilde{q}} >  8.0 \times 10^{-9}$ is required in order to decay the neutralino within 1 mm. 

For the heavier neutralino $\tilde{\chi}^0_2$ and the chargino $\tilde{\chi}^\pm$ in the Higgsino LSP case, both particles are assumed to be dominated by the RPC decay, {\it i.e.}, $\tilde{\chi}^0_2 \to h^*/Z^* \tilde{\chi}^0_1$ and $\tilde{\chi}^\pm \to W^* \tilde{\chi}^0_1$. Because of the compressed spectrum of the Higgsino sector, the final states from the off-shell bosons ($h^*, Z^*, W^*$) are too soft to be detected and only the $\tilde{\chi}^0_1$ is visible. 
{As a result, each of the three Higgsinos ($\tilde{H}^\pm, \tilde{H}^0_{1,2}$) perform as three jets at the detector with one of the jets being $b$-tagged.}
In fact, if the RPV decays of $\tilde{\chi}^0_2$ and $\tilde{\chi}^\pm$ are important, {\it i.e.}, the $\tilde{\chi}^0_2 \to b c s$ and $\tilde{\chi}^\pm \to t c s/ s s b/ c c b$ are open. The detector signals remain the same, except for the $\tilde{\chi}^\pm \to t c s$ channel, which produces an extra top quark in the final state.

\section{Testing with LHC searches} \label{sec:lhc}

In this work, our signal events are generated by \madgraph~\cite{Alwall:2014hca}, where \pythia~\cite{Sjostrand_2014zea}, \fastjet~\cite{Cacciari_2011ma} and \delphes~\cite{deFavereau_2013fsa} have been used to implement parton showering, jet reconstruction and detector effects.

\subsection{Analysis and results under $13$ TeV data} \label{sec:lhc13}
As discussed in the previous section, our signals include dijet resonances pair ($\tilde{b} \to cs$), multijet ($\tilde{t} \to b \tilde{\chi}^\pm$/$\tilde{b} \to b \tilde{\chi}^0$) and lepton + jets ($\tilde{t} \to t \tilde{\chi}^0$/$\tilde{b} \to t \tilde{\chi}^\pm$). Even though most of our specific signals have not been searched at the LHC yet, there are some existing searches for the similar final states which could already constrain our signal processes. 
We will recast three relevant RPV SUSY searches from ATLAS: a lepton plus high jet multiplicity search~\cite{Aaboud_2017faq}, pair-produced resonances in four-jet final states~\cite{Aaboud_2017nmi}, and multijet final states~\cite{ATLAS-CONF-2016-057}. The validations of our recasting are provided in appendixes. 

To derive the bounds from recasting, a variable $R^{ai} = N^{ai}_{\text{NP}} / N^{ai}_{UL}$ is defined in each signal region $i$ of each analysis $a$, where $N^{ai}_{\text{NP}} $ is the number of our signal events in the signal region $i$ of analysis $a$ obtained from our simulation and $N^{ai}_{UL}$ is the observed 95\% CL model independent upper limit provided in each experimental analysis. The maxima $R^{\max} = \max_{a,i} \{R^{ai} \}$ is defined as the most sensitive one from all of the searches. This means a signal point is excluded by the current search if $R^{\max} >1$. 

\begin{figure}[thb]
\begin{center}
\includegraphics[width=0.45\textwidth]{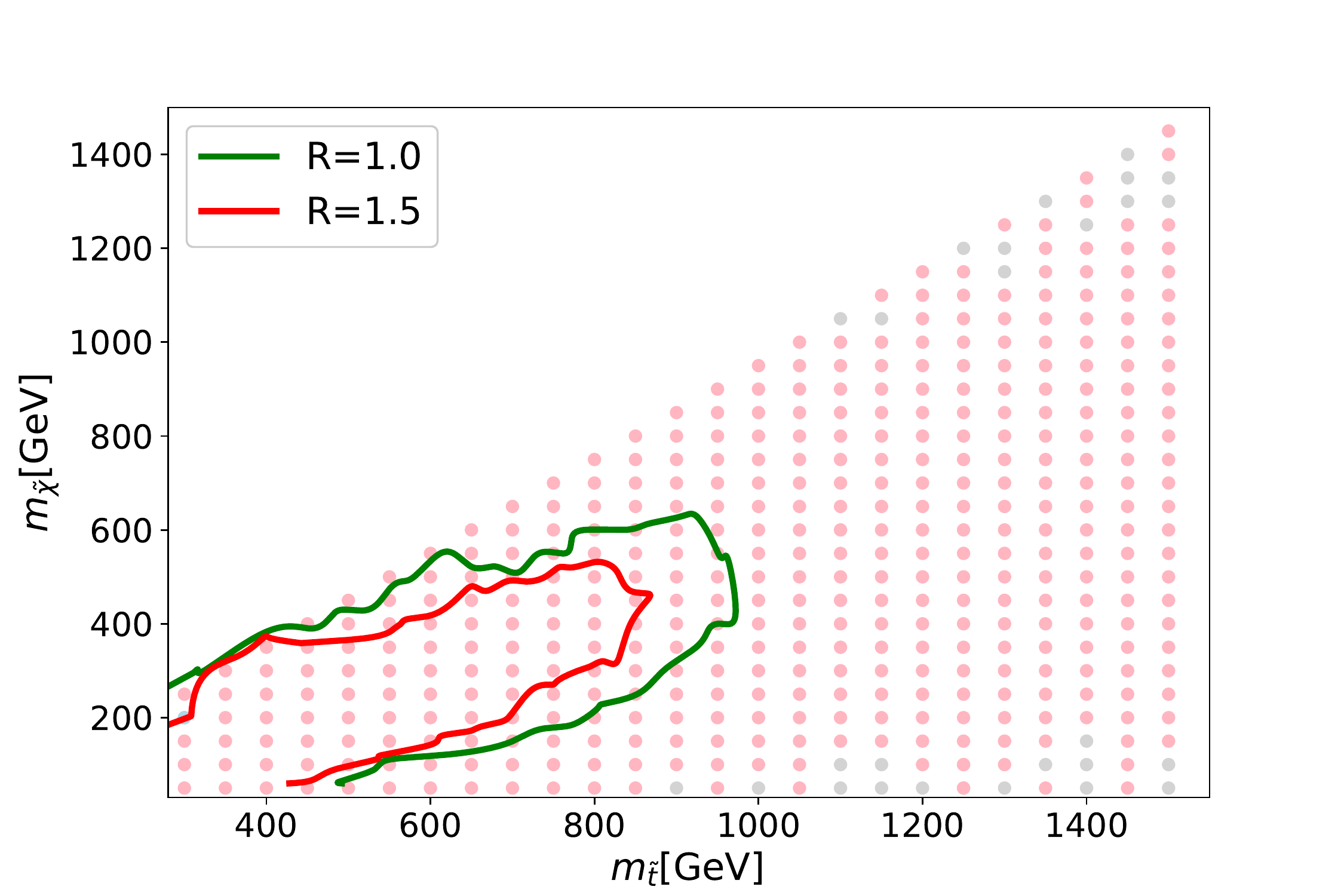}
\includegraphics[width=0.45\textwidth]{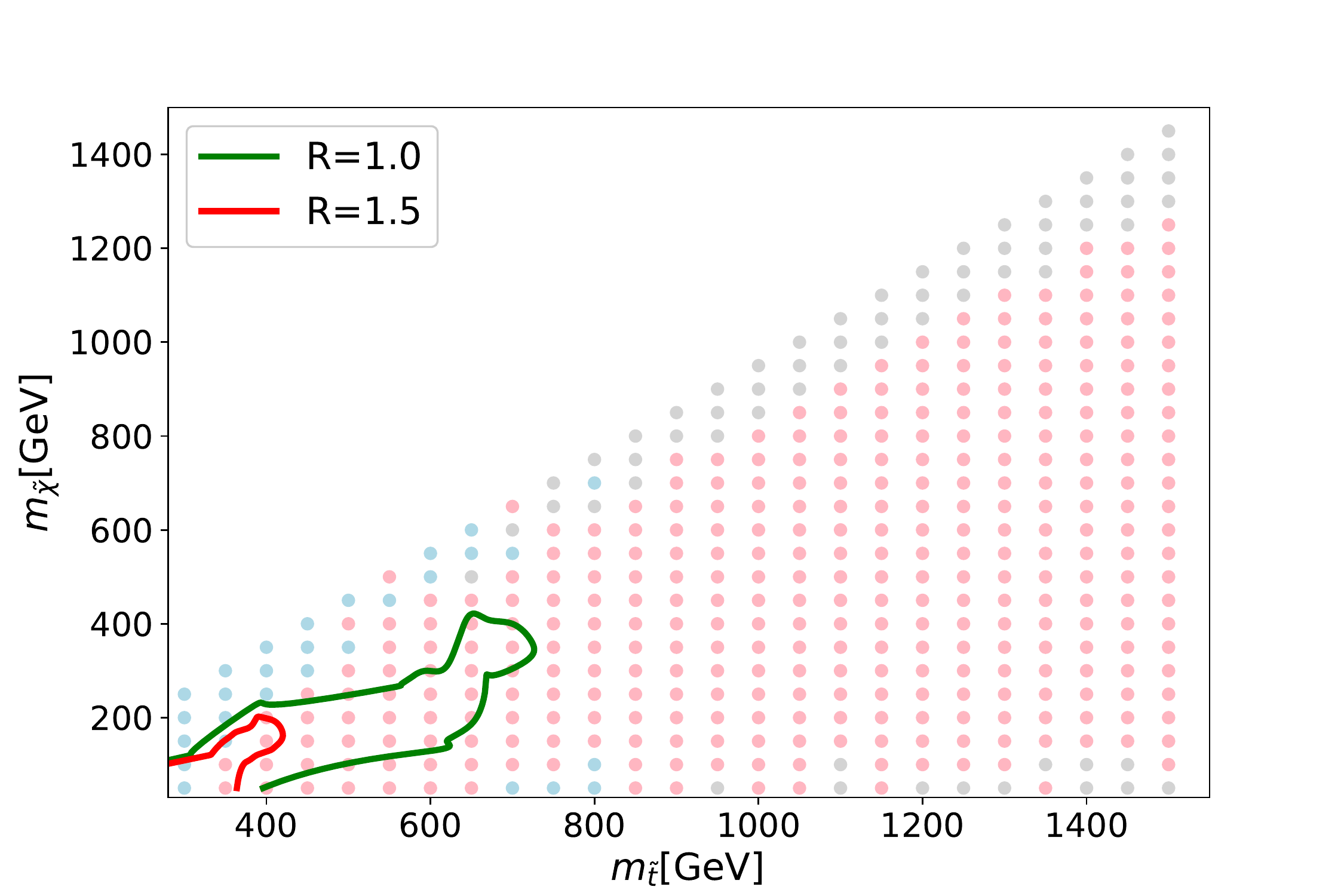}
\end{center}
\caption{\label{fig:lhcstop} Left panel: bounds on the $\tilde{t} - \tilde{B}$ simplified model. Right panel: bounds on the $\tilde{t} - \tilde{H}$ simplified model. The green and red contours correspond to exclusion limits with $R^{\max}=1.0$ and $R^{\max}=1.5$.  
The most sensitive analysis at each grid is indicated by the point colors.  Points with the colors of pink, blue and grey correspond to the analyses in Refs.~\cite{Aaboud_2017faq}, \cite{Aaboud_2017nmi}, 
and \cite{ATLAS-CONF-2016-057}, respectively. }
\end{figure}

In Fig.~\ref{fig:lhcstop}, we plot the contours of $R^{\max}=1.0$ and $R^{\max}=1.5$ on the $m_{\tilde{t}}$-$m_{\tilde{\chi}^0}$ plane for the $\tilde{t} - \tilde{B}$ and $\tilde{t} - \tilde{H}$ simplified model. 
The most sensitive search on each grid is indicated by the point color: pink, blue and grey points corresponding to the lepton plus high jet multiplicity analysis~\cite{Aaboud_2017faq}, pair-produced resonances in the four-jet final state analysis~\cite{Aaboud_2017nmi} and multijet final state analysis~\cite{ATLAS-CONF-2016-057}, respectively.

In the $\tilde{t} - \tilde{B}$ simplified model, because of the on-shell/off-shell top quark in the final state which could decay leptonically, the lepton plus jets search is the most sensitive one for most of the time. The constraint on this model is quite stringent, except for the regions with $m_{\tilde{t}} \sim m_{\tilde{\chi}^0}$ or relatively light bino. In the former region the lepton from the top decay is too soft. While for a too light bino, the three jets from a RPV $\tilde{\chi}^0$ decay will be collimated, performing as a single jet in the detector. Thus, the jet multiplicity in the final state is reduced. 

The bounds obtained in the $\tilde{t}-\tilde{H}$ simplified model are much weaker, mainly because of the branching ratio suppression for each channel; {\it i.e.}, $\tilde{t} \to t \tilde{\chi}^0$ with $t \to b W(\to \ell \nu)$ and $\tilde{t} \to b \tilde{\chi}^\pm$ produce 
the final states with and without a detectable lepton. From the figure, we can see that still the lepton plus jets search is the most sensitive one in most regions. This means that the current searches are only sensitive to the $\tilde{t} \to t \tilde{\chi}^0$ while the $\tilde{t} \to b \tilde{\chi}^\pm$ mode which produces an energetic $b$-jet is overlooked. 

\begin{figure}[thb]
\begin{center}
\includegraphics[width=0.45\textwidth]{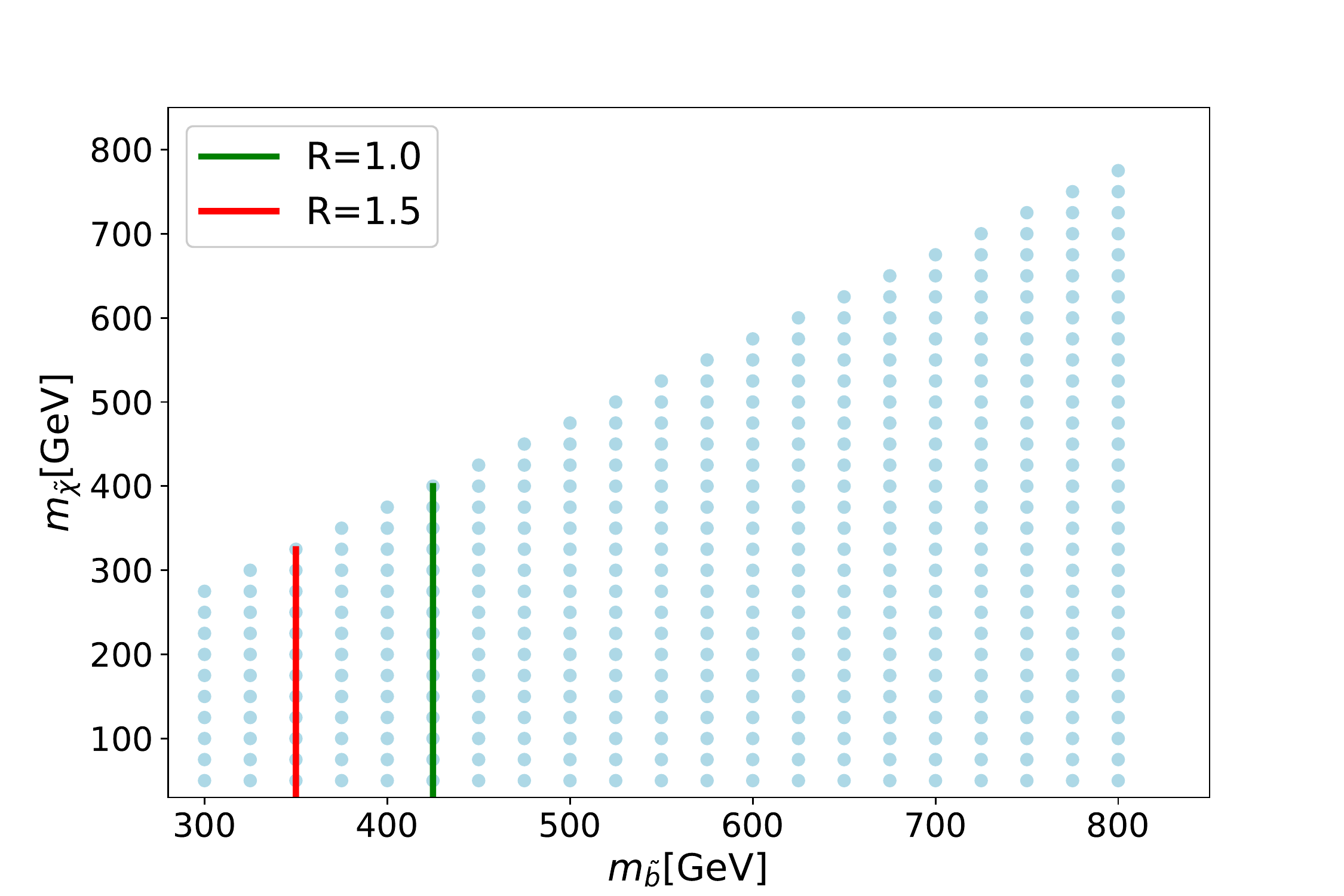}
\includegraphics[width=0.45\textwidth]{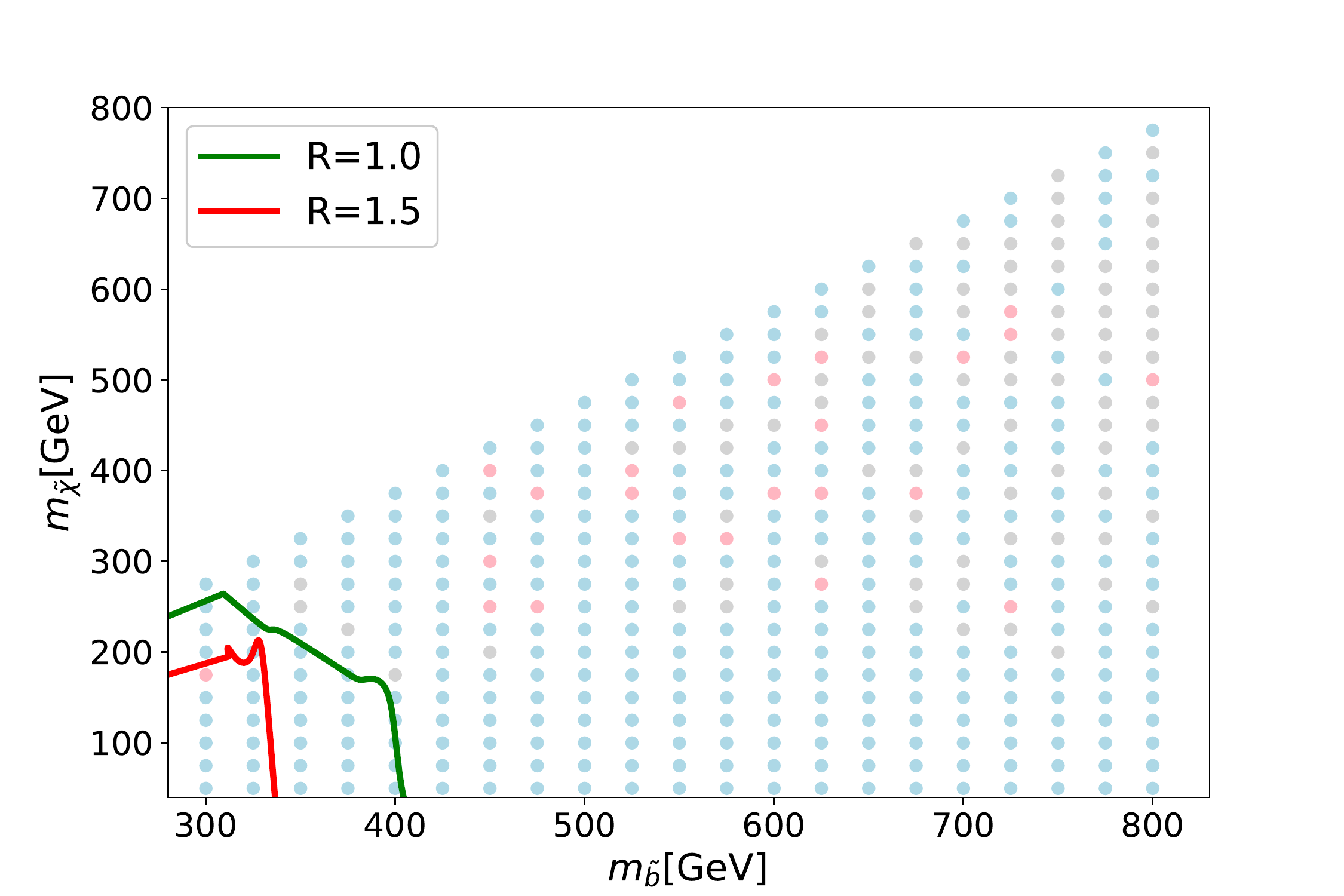}
\end{center}
\caption{\label{fig:lhcbb1} Bounds on the $\tilde{b}-\tilde{B}$ simplified model with $\text{Br}(\tilde{b} \to cs) =100\%$ (left) and $\text{Br}(\tilde{b} \to b \tilde{\chi}^0) =100\%$ (right).  The lines and point styles are same with Fig.~\ref{fig:lhcstop}.}
\end{figure}

For both the $\tilde{b}-\tilde{B}$ and $\tilde{b}-\tilde{H}$ simplified models, there is a direct RPV sbottom decay $\tilde{b} \to  cs$. At the LHC, there is a search~\cite{Aaboud_2017nmi} for a stop pair which decays into $s d$ or $b s$ through nonzero $\lambda''_{312}$ or $\lambda''_{323}$ which coincide with our scenarios when $\text{Br}(\tilde{b} \to cs) =100\%$. The corresponding bounds are presented in the left panel of Fig.~\ref{fig:lhcbb1}. Similar to the stop case, the sbottom with a mass below $\sim 425$ GeV has been excluded in this scenario.

In the $\tilde{b}-\tilde{B}$ simplified model, besides the direct RPV decay, an sbottom can decay into $b \tilde{\chi}^0$ with a subsequent RPV decay $\tilde{\chi}^0 \to b c s$. This channel is the most difficult channel with respect to current searches: 1) It does not produce any lepton in the final state. 2) For $m_{\tilde{b}} \sim 300$-$400$ GeV, the final state jets are typically too soft to pass the jet selections in the multijet search. As we can see from the right panel of Fig.~\ref{fig:lhcbb1}, the LHC searches are only able to exclude the corner with both the light sbottom and neutralino. In this region, the sbottom cross section is large and the neutralino is reconstructed as a single jet because of its collimated decay products.
So the signal here appears to be similar as the dijet resonance $\tilde{b} \to j j$. The four-jet resonances search~\cite{Aaboud_2017nmi} provides the most sensitive probing in most regions. Especially, for a very light neutralino $m_{\tilde{\chi}} \sim 25$ GeV, the sbottom is excluded up to 425 GeV in this model, which is close to the limit obtained in the $\text{Br}(\tilde{b} \to cs) =100\%$ scenario. 

We have also performed the test on the scenario with $\text{Br}(\tilde{b} \to cs)=\text{Br}(\tilde{b} \to b \tilde{\chi}^0) = 50\%$. 
Because both channels are dominantly constrained by the same search, {\it i.e.}, the dijet resonances search~\cite{Aaboud_2017nmi}, we find the distributions of $R^{\max}$ of this scenario is similar to the right panel of Fig.~\ref{fig:lhcbb1}. Note that the rate of true dijet events is reduced to 25\%.

\begin{figure}[thb]
\begin{center}
\includegraphics[width=0.45\textwidth]{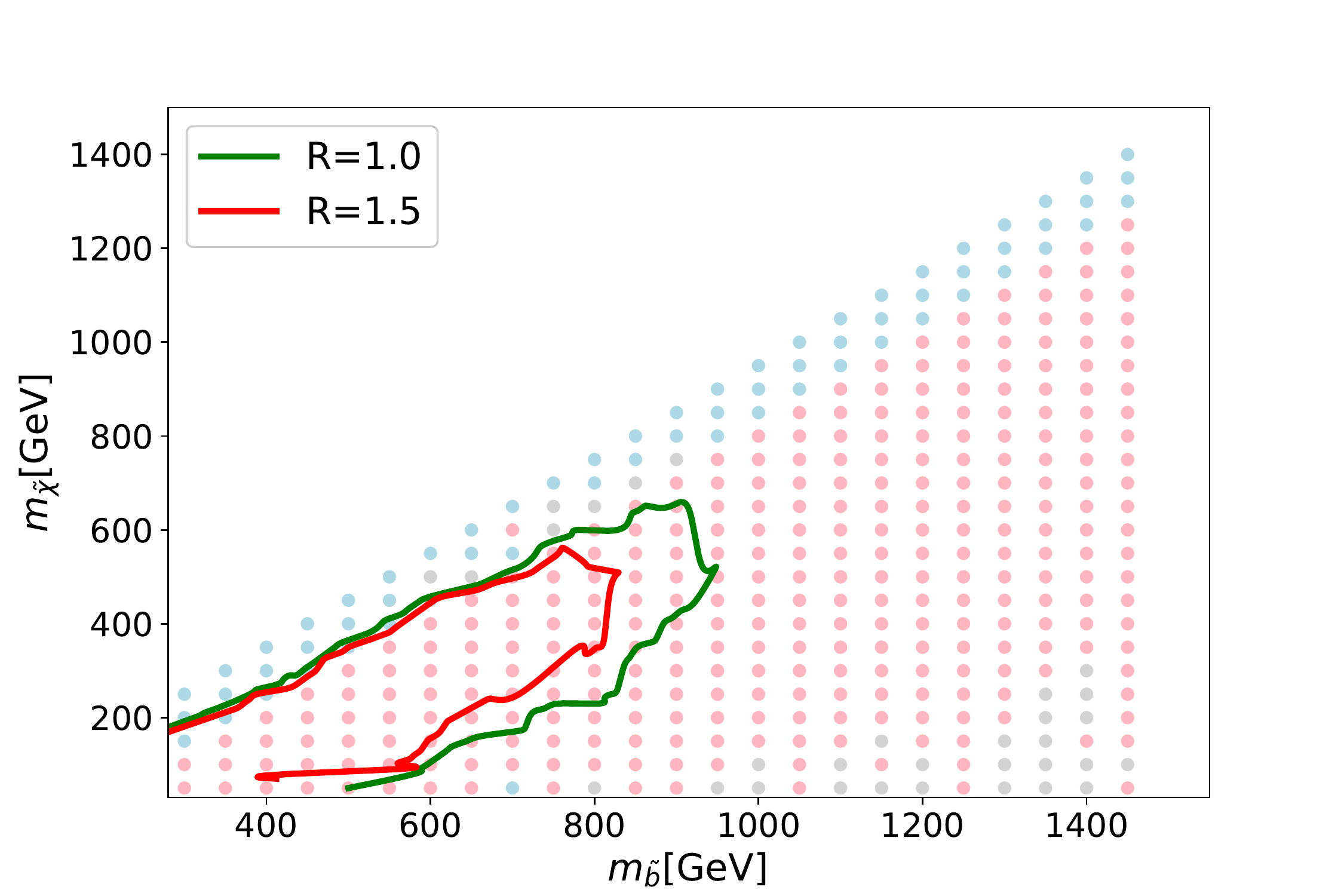}
\includegraphics[width=0.45\textwidth]{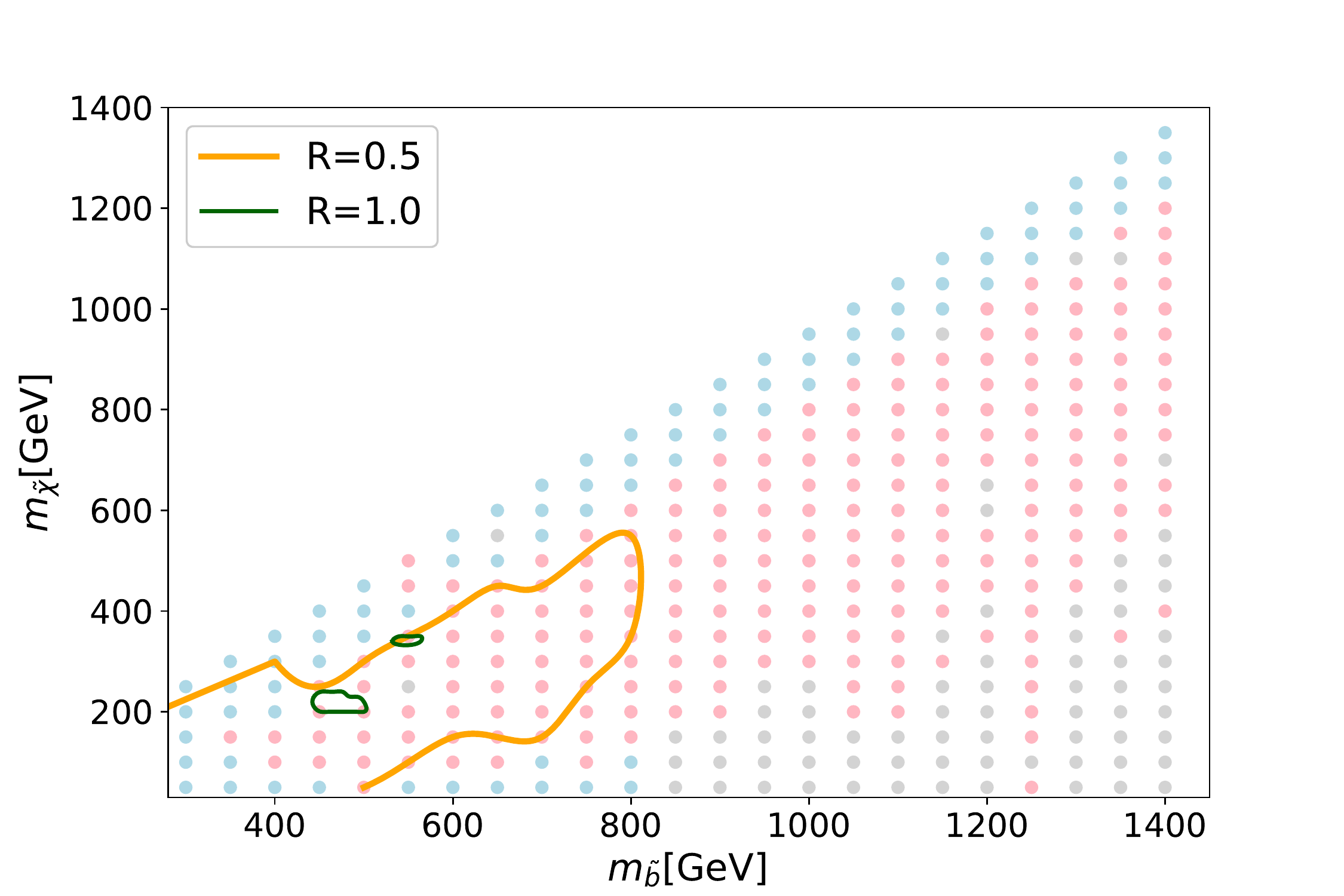}
\end{center}
\caption{\label{fig:lhcsbh} Bounds on the $\tilde{b}-\tilde{H}$ simplified model with $\text{Br}(\tilde{b} \to cs) =0\%$ (left) and  $\text{Br}(\tilde{b} \to cs) =50\%$ (right).  The lines and point styles are the same as Fig.~\ref{fig:lhcstop}. In the right panel, $R^{\max}$ values are always less than 1.5, so the contours of $R^{\max}=0.5$ and $R^{\max}=1.0$ are presented. }
\end{figure}

Finally, for the $\tilde{b}-\tilde{H}$ simplified model, if the direct RPV decay of sbottom is subdominating, its final states are similar with that of the $\tilde{t}-\tilde{H}$ simplified model for pure right-handed sbottom and similar with that of $\tilde{t}-\tilde{B}$ simplified model when there is a small component of a left-handed sbottom. Taking the later case as an example,  the left-handed sbottom mixing is taken to be 0.1 so that $\tilde{b} \to t \tilde{H}^\pm$ dominates. The bounds are shown in the left panel of Fig.~\ref{fig:lhcsbh} which is slightly weaker than that in the left panel of Fig.~\ref{fig:lhcstop}, due to the branching ratio suppression. 
The lepton plus jets search is the most sensitive one, which excludes the region with $m_{\tilde{b}} - 500~\text{GeV} \lesssim  m_{\tilde{H}} \lesssim m_{\tilde{b}}-m_{t}$. 
The scenario with comparable branching ratios of direct RPV decay $\tilde{b} \to  c s$ and $\tilde{b} \to t \tilde{H}^\pm/b\tilde{H}^0$ decay will be more difficult to probe, due to further branching ratio suppression. The corresponding bounds with $\text{Br}(\tilde{b} \to cs) =50\%$ are shown in the right panel of Fig~\ref{fig:lhcsbh}. It shows that the current search can only exclude the region with $m_{\tilde{b}} \sim [400,500]$ GeV and $m_{\tilde{H}}\sim 200$ GeV.

\subsection{Prospects with higher luminosity}

From our above study, we have shown that the current searches are not yet able to exclude most of the parameter space, especially in the $\tilde{t}-\tilde{H}$ simplified model, the $\tilde{b}-\tilde{B}$ simplified model, and the $\text{Br}(\tilde{b} \to cs) =50\%$ scenario in the $\tilde{b}-\tilde{H}$ simplified model. However, the $R^{\max}$ values on the most of the grids in those scenarios are already around $\mathcal{O}(0.1)$. 
It will be interesting to see the prospects of the sensitivity at higher luminosity LHC. In the following, we will simply extrapolate the exclusion limits at current stage to that of the future 14 TeV LHC with an integrated luminosity of 3000 fb$^{-1}$.

The following assumptions as adopted in Ref.~\cite{CMS_2013xfa} are made
\begin{itemize}
\item The definitions of signal regions remained the same. Moreover, for both signal and background events, the selection efficiencies of each signal region are almost kept the same from 13 TeV to 14 TeV. 
\item The statistical uncertainty of the background is rescaled by $\sqrt{B}$, where $B$ is the total number of background events in the most sensitive signal region, {\it i.e.}, the one that provides $R^{\max}$. 
\item The systematic uncertainty of the background is proportional to the $B$. 
According to the analyses in Refs.~\cite{Aaboud_2017faq,Aaboud_2017nmi, ATLAS-CONF-2016-057}, the systematic uncertainties in the numbers of background events of signal regions are always less than $\sim 10\%$. In most cases, they are less than 5\%. We will take the systematic uncertainty to be 5\% in the extrapolation~\footnote{We have tried to plot the exclusion contours for 14 TeV prospects by taking the systematic uncertainty in each signal region to be 20\%. Because of the sizeable background uncertainty, the improvements of exclusion bounds at the future LHC is tiny compared to those in the existing LHC analyses.}. 
\item In addition, we assume the observed total number of events in each signal region to be the same with the background expectation.
\end{itemize}

\begin{figure}
\centering
\subfigure[$\tilde{t}-\tilde{H}$]{
\label{Fig.sub.1}
\includegraphics[scale=0.35]{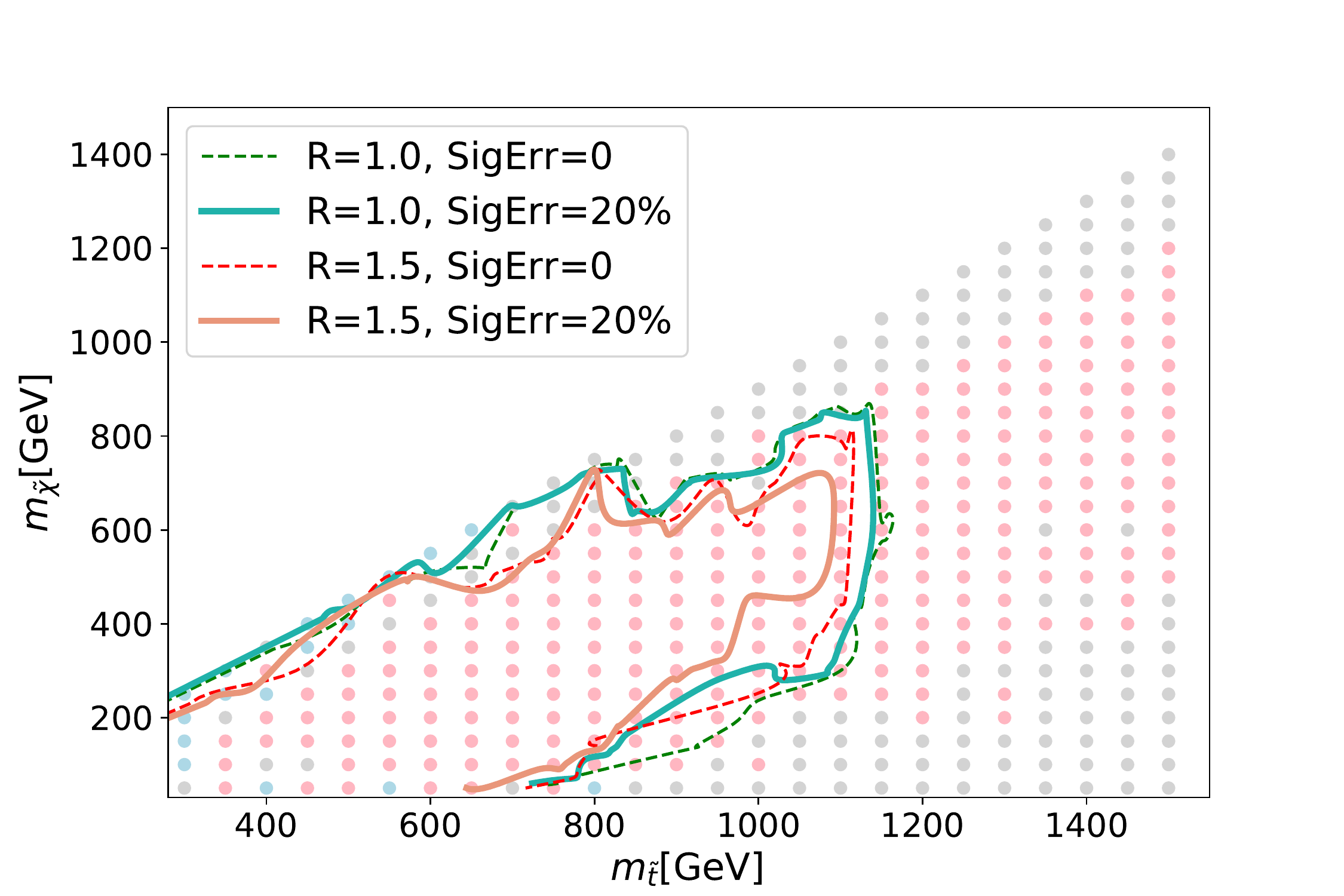}}
\subfigure[$\tilde{b} \to cs$]{
\label{Fig.sub.2}
\includegraphics[scale=0.35]{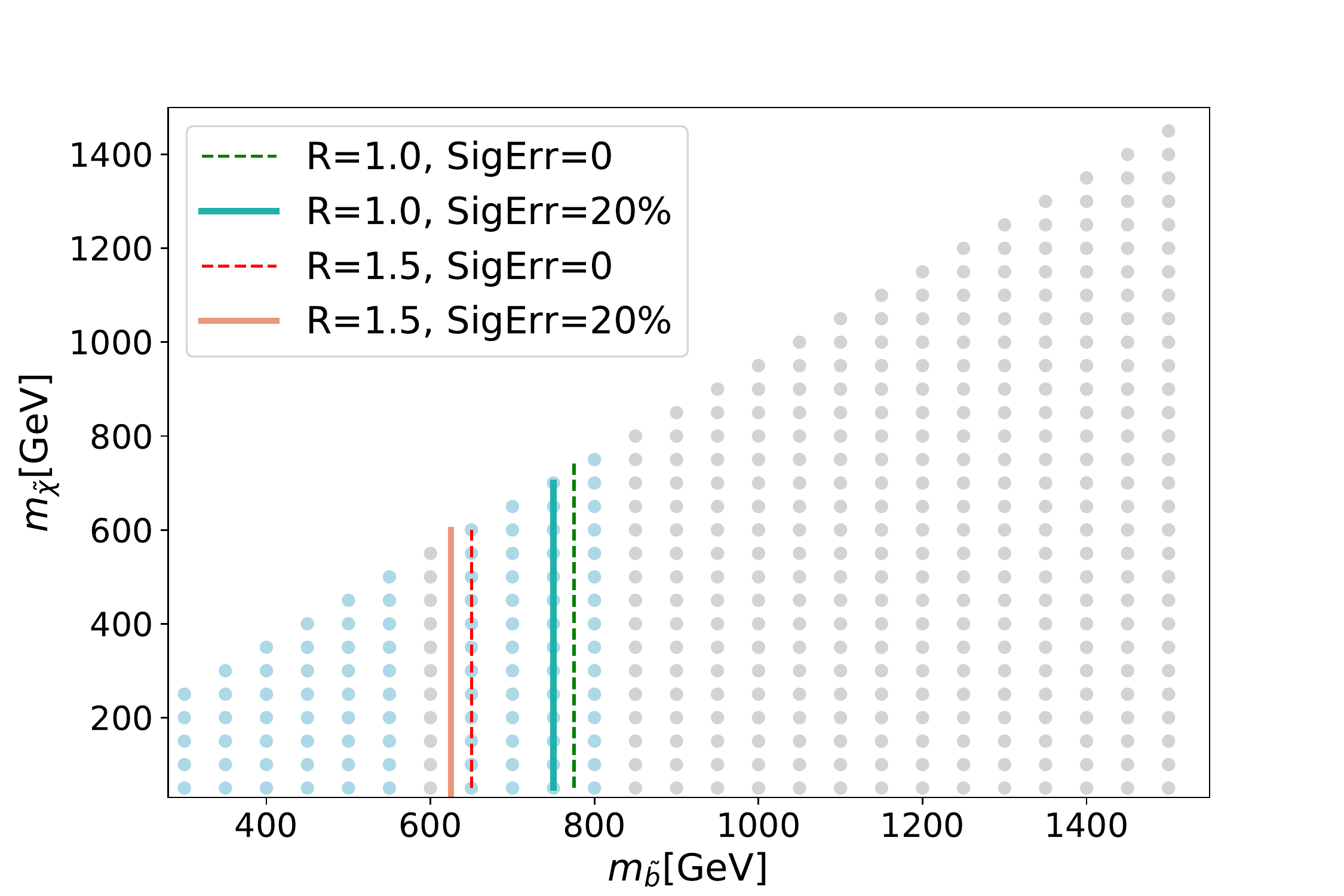}}
\subfigure[$\tilde{b}-\tilde{B}$ with $\text{Br}(\tilde{b} \to cs) =0\%$]{
\label{Fig.sub.3}
\includegraphics[scale=0.35]{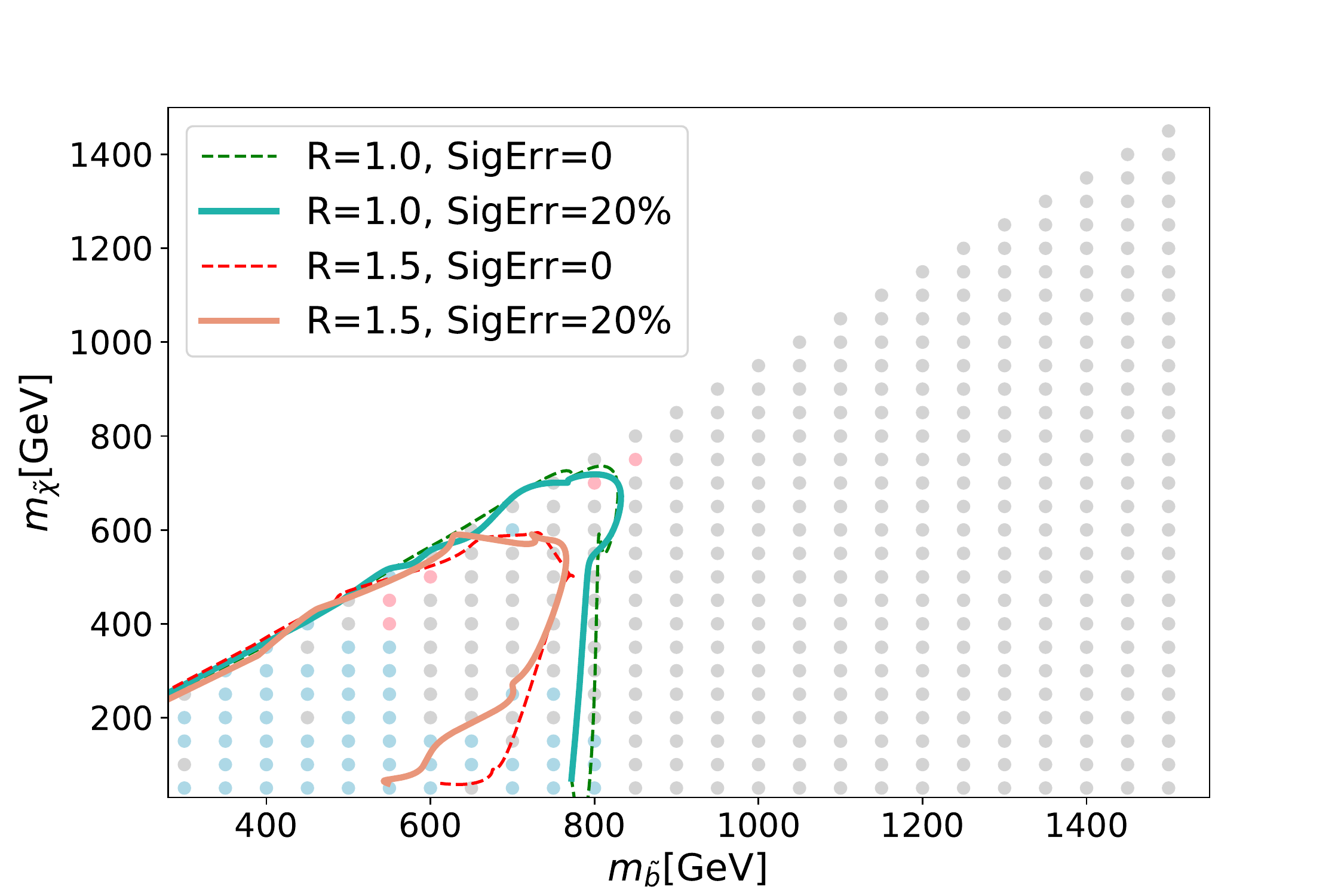}}
\subfigure[$\tilde{b}-\tilde{H}$ with $\text{Br}(\tilde{b} \to cs) =50\%$]{
\label{Fig.sub.4}
\includegraphics[scale=0.35]{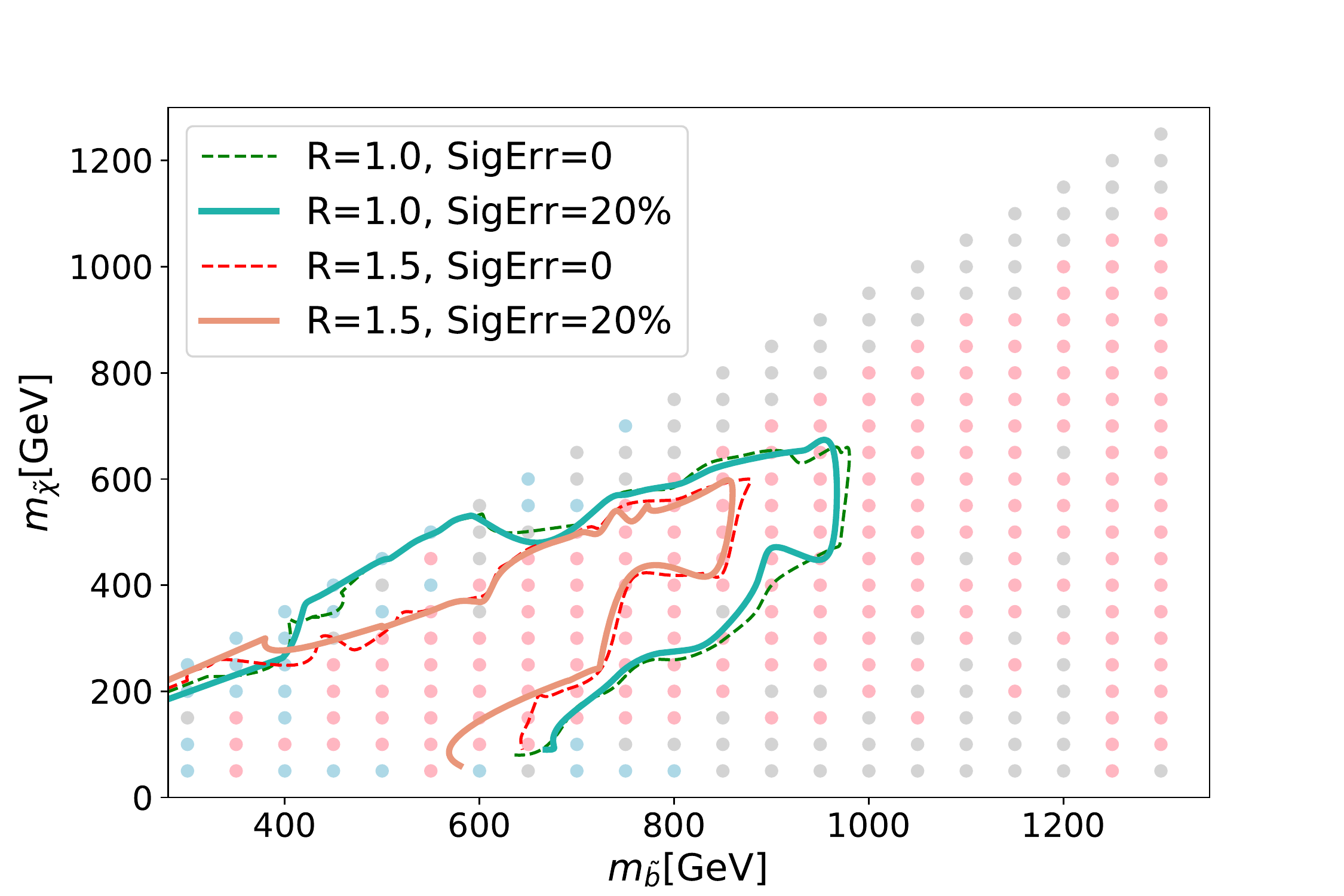}}
\caption{Expected signal reaches at 14 TeV LHC with integrated luminosity of 3000 fb$^{-1}$. The exclusion limits for four cases that are weakly constrained by the current LHC searches are given. 
The lines and point styles are same as Fig.~\ref{fig:lhcstop}.
In addition, the corresponding bounds with and without the signal uncertainty of 20\% are indicated by solid and dashed line, respectively.  \label{ext14}}
\end{figure}

With these assumptions, the total number of signal and background events in a signal region is rescaled by a factor of 
\begin{equation}
F_{\text{sig(bkg)}}=\frac{\mathscr{L}_0}{\mathscr{L}'} \times \frac{\sigma_{\text{sig (bkg)}}^{14}}{\sigma_{\text{sig (bkg)}}^{13}},
\end{equation}
where $\sigma_{\text{sig(bkg)}}^{13 (14)}$ is the production cross section of a signal or background process at 13 TeV or 14 TeV. $\mathscr{L}'$ and $\mathscr{L}_0$ are the integrated luminosities at 13 TeV and 14TeV, respectively. Note only the dominant background process in each analysis is considered to estimate the scaling of background cross section. That is $t\bar{t}+$jets for the analysis in Ref.~\cite{Aaboud_2017faq} and QCD multijets for the analyses in Refs.~\cite{Aaboud_2017nmi, ATLAS-CONF-2016-057} respectively.
In the absence of any systematic errors, given the rescaled total number of background ($N_b$) and signal ($N_s$) events, the probability for observing $N_b$ events with an expected mean number of events $\mu= N_s +N_b$ follows the Poisson distribution or Gaussian distribution~\cite{CDF_sta}
\begin{equation}
P(N_{b};\mu) = 
\begin{cases}
   \frac{\mu^{N_{b}}e^{-\mu}}{N_{b}!},  ~~~~\text{for } N_{b} \leq 100 \\
   \frac{e^{\frac{(N_b-\mu)^2}{2\mu}}}{\sqrt{2\pi\mu}},~~~~ \text{for } N_{b}>100.
\end{cases}
\end{equation}
The effects of systematic uncertainties of background ($\sigma_b$) and signal ($\sigma_s$) can be accommodated by convoluting the probability with Gaussian function that is representing the prior probability density of each parameter~\cite{Junk:1999kv}. This gives the likelihood as
\begin{align}\label{uplimit2}
\mathcal{L}(N_{b}|N_s,N_b,\sigma_b, \sigma_s)=\frac{1}{2\pi\sigma_s\sigma_b}\int_{-5\sigma_s}^{5\sigma_s}d\delta_s\int_{-5\sigma_b}^{5\sigma_b}d\delta_bP(N_{b};\mu)e^{\frac{\delta_b^2}{2\sigma_b^2}}e^{\frac{\delta_s^2}{2\sigma_s^2}}~.~
\end{align}
Having the likelihood, Bayes's Theorem~\cite{doi:10.1119/1.17901} can be used to derive a posterior probability for any signal events number $S$
\begin{align}
\mathcal{P}(S|N_s,N_b,\sigma_b, \sigma_s) = \frac{\mathcal{L}(N_b|S,N_b,\sigma_b)P(S)}{\int_{0}^{\infty}\mathcal{L}(N_b|S',N_b,\sigma_b)P(S') d S'}~,~
\end{align}
where $P(S)$ is the prior probability of signal event number which is assumed to be uniform for all $S>0$. 
The 95\% CL upper limit on the signal event number $N_{\text{limit}}$ can be evaluated by
\begin{align}\label{uplimit1}
\int_0^{N_{\text{limit}}} \mathcal{P}(S|N_s,N_b,\sigma_b, \sigma_s) dS = 0.95. 
\end{align}
Finally, in each signal region $i$, the ratio ($R^i_{14}$) between the rescaled number of signal events $N_s$ and the $N_{limit}$ is calculated. The maximal ratio $R^{\max}_{14} = \max_i\{R^i_{14}\}$ among all signal regions is used to test a given model.

The extrapolated exclusion bounds for these difficult scenarios are shown in Fig.~\ref{ext14}~\footnote{In the $\tilde{t} - \tilde{B}$ simplified model and $\tilde{b}-\tilde{H}$ simplified model with $\text{Br}(\tilde{b} \to cs) =0\%$, the final state is mostly $t\bar{t} + (jjj) + (jjj)$. The 14 TeV LHC with integrate luminosity of 3000 fb$^{-1}$ will be able to reach stop/sbottom mass up to 1.5 TeV. Ref.~\cite{Duggan:2013yna} studied the same channel and gave a relatively stronger bound, i.e. $m_{\tilde{t}} \lesssim 1.7$ TeV can be excluded. }. 
Because of the increased signal event number and smaller systematics that we have assumed at 14 TeV with an integrated luminosity of 3000 fb$^{-1}$, the branching ratio suppression in the $\tilde{t}-\tilde{H}$ simplified model becomes a less severe problem. There will be sufficient events with leptonic final states for most of the points with $m_{\tilde{t}} \lesssim 1$ TeV. So the search for a lepton plus high jet multiplicity excludes most of the regions, except those with relatively degenerate spectra so that the lepton is too soft to be detected and those with heavy stop and light neutralino so that jet multiplicity is low. 
The upper-right panel of Fig.~\ref{ext14} shows the exclusion limits for the sbottom pair production followed by direct RPV decay $\tilde{b} \to cs $. The extrapolated dijet pair resonances search will push the bounds on sbottom mass to $\sim 750$-$800$ GeV, depending on the signal uncertainty~\footnote{This limit is weaker than the one obtained in Ref.~\cite{Duggan:2013yna}, which shows that the future LHC can reach the stop mass up to $\sim$1 TeV in this channel.}. 
In the higher mass region, where both the number and the energies of initial state radiated jets are increased, the multijet search becomes the most sensitive. 
As has been discussed in Sec.~\ref{sec:lhc13}, the $\tilde{b}-\tilde{B}$ is the most difficult model with respect to current searches. Moreover, including the direct RPV decay of sbottom does not change the current sensitivities on this model. In the lower-left panel of Fig.~\ref{ext14}, we present the $R_{14}^{\max}$ distribution in the scenario with $\text{Br}(\tilde{b} \to cs) =0\%$. (We verified that the scenario with $\text{Br}(\tilde{b} \to cs) =50\%$ gives the similar result.) 
The future prospects for this model are much more promising. 
The lower sbottom mass region is constrained by the dijet resonant search, while the multijet search provide the strongest constraint in the high sbottom mass region. The future LHC can reach the sbottom mass up to $\sim 800$ GeV. 
The search sensitivity to the $\tilde{b}-\tilde{H}$ model is much better in some mass regions, where the leptons in the final state can be energetic. For our choice of sbottom mixing, the signature of $\text{Br}(\tilde{b} \to cs) =0\%$ scenario of the $\tilde{b}-\tilde{H}$ simplified model is similar to that of the $\tilde{t}-\tilde{B}$ simplified model, as has been found in the recasting, i.e., the left panels of Fig.~\ref{fig:lhcstop} and Fig.~\ref{fig:lhcsbh}. In the lower-right panel of Fig.~\ref{ext14}, we present the bound for the $\text{Br}(\tilde{b} \to cs) =50\%$ scenario of this model; the constraint of which is weaker than that of the $\text{Br}(\tilde{b} \to cs) =0\%$ scenario, simply because of the branching ratio suppression. The lepton plus jet search can exclude the sbottom mass in this scenario up to $\sim 950$ GeV.

\section{Conclusion}  \label{sec:conclude}
In this paper, we proposed a RPV SUSY scenario that is the least constrained by current LHC searches and low energy experiments, in which only the $U^c_2 D^c_2 D^c_3$ operator is nonzero. Motivated by the naturalness argument, four simplified models with a relatively light stop/sbottom are considered, {\it i.e.}, $\tilde{t} - \tilde{B}$, $\tilde{t} - \tilde{H}$, $\tilde{b}-\tilde{B}$, and $\tilde{b} - \tilde{H}$ models. 

Those simplified models can lead to collider signatures of multiple jets, dijet pair resonances as well as leptons plus jets if any on-shell/off-shell top quarks are produced. By recasting the relevant LHC searches onto our simplified model, we found some difficult scenarios regarding the current searches, where the stop/sbottom masses are barely constrained. They are $\tilde{t}-\tilde{H}$ simplified model, the $\tilde{b}-\tilde{B}$ simplified model, and the $\text{Br}(\tilde{b} \to cs) =50\%$ scenario in the $\tilde{b}-\tilde{H}$ simplified model.
Next, we extrapolated those existing searches to higher energy and luminosity LHC, {\it i.e.},
the 14 TeV 3000 fb$^{-1}$ LHC. Under our assumptions in the extrapolation, the future prospects of the LHC sensitivities to those difficult scenarios are promising. Especially, the stop/sbottom up to 1.1 TeV can be probed in the $\tilde{t}-\tilde{H}$ simplified model. 
In the $\tilde{b}-\tilde{B}$ simplified model with either $\text{Br}(\tilde{b} \to cs) =0\%$ or 100\%, the sbottom mass can be reached up to $\sim 800$ GeV. Note that the signature of the $\tilde{b}-\tilde{B}$ simplified model with $\text{Br}(\tilde{b} \to cs) =0\%$ is featured by four b-jets and each two of them have the same origin, all current searches are not optimized for it. We expected that an improved search,
 which utilizes these special features, can be more sensitive to this model.   
As for the $\text{Br}(\tilde{b} \to cs) =50\%$ scenario in the $\tilde{b}-\tilde{H}$ simplified model, the branching ratio suppression is substantial, and the sbottom with mass $\sim 600$ GeV is still safe.

\section*{Acknowledgements}
We thank Chuang Li and Yizhou Fan for useful discussions. We especially thank Angelo Monteux for helping us to prove some of the results.
In addition, we thank the Korea Institute for Advanced Study for providing computing resources (KIAS Center for Advanced Computation Linux Cluster System) for this work.

This research was supported in part by the National Natural Science Foundation of China under grants No. 11475238 and No. 11647601, by Key Research Program of Frontier Science, CAS and {by the Fundamental Research Funds for the central Universities.}
The numerical results described in this paper have been obtained via the HPC Cluster of ITP-CAS.

\begin{appendix} 
\section{Search for a lepton plus multijet final states} \label{app1}
Reference~\cite{Aaboud_2017faq} searches for a final state with multijets and a lepton at 13 TeV 36 fb$^{-1}$ LHC. To validate our recast, we take the second model in the paper, {\it i.e.}, $ p p \to \tilde{g} \tilde{g} \to \bar{t} \tilde{t}(\to \bar{b} \bar{s})  \bar{t} \tilde{t}(\to \bar{b} \bar{s})$. 

We generate events for $p p \to \tilde{g} \tilde{g}$ with \madgraph interfaced to \pythia, which is used to decay $\tilde{g}$.  In this step, the parton distribution function is provided by the \nnpdf. Then we simulate the detector effects via \delphes, including pileup effects. The mass and detector parameters are $M_{\tilde{t}}=1.0 $~TeV, 
$M_{\tilde{g}}=1.6 $~TeV, and the $b$-tag efficiency is 78\%.

In this analysis, only the total numbers of the background and their uncertainties as well as the observed event numbers are given (we will take the data in Table~II of Ref.~\cite{Aaboud_2017faq} as an example). We can use the Eq.~(\ref{uplimit1}) to calculate the new physics upper limit for each signal region.  The uncertainty of the signal event number $\sigma_s$ is assumed to be $\sigma_s=0.1N_s$. Our results are given in Table~\ref{tab:bench1}.

\begin{table}[!h]\centering
 \begin{tabular}{|c|c|c|c|c|c|c|} \hline
  \multicolumn{1}{|c|}{\multirow{1}{*}{    }}
   &\multicolumn{2}{|c}{\multirow{1}{*}{$\geq 10 jets$}}
    &\multicolumn{2}{|c|}{\multirow{1}{*}{$\geq 11 jets$}}
     &\multicolumn{2}{c|}{\multirow{1}{*}{$\geq 12 jets$} } \\ \hline
     
  \multicolumn{1}{|c|}{\multirow{2}{*}{Bkg}}
   & 0b & $\geq 3b$ & 0b & $\geq 3b$ & 0b & $\geq 3b$  \\  \cline{2-7}   
   & $26 \pm 4$ & $60 \pm 6$ & $4.5 \pm 1.0$ & $12.6 \pm 1.9$ & $0.87 \pm 0.23$ & $2.5 \pm 0.7$ \\ \hline
    Data & 23 & 61 & 5 & 16 & 0 & 4 \\ \hline
    Upper limit & 12.7 & 24.5 & 7.0 & 13.2 & 3.0 & 7.1 \\ \hline
 \end{tabular}
  \caption{\label{tab:bench1} The background and data are taken from Table~II of Ref.~\cite{Aaboud_2017faq}. The last row gives the 95\% CL new physics upper limit.}
\end{table}

The final selection efficiencies of all signal regions for our benchmark point have been given in the auxiliary Table~II of Ref.~\cite{Aaboud_2017faq}, which is referred in the row of ``Exp" in our Table~\ref{tab:bench7}. For comparison, the corresponding efficiencies from our simulation and from CheckMATE~\cite{Dercks:2016npn} are provided in the third and fourth row of the same table, denoted as ``Sim" and ``Sim2", respectively.

\begin{table}[!h]\centering
 \begin{tabular}{|c|c|c|c|c|c|c|}\hline
  \multicolumn{1}{|c|}{}
   & $60\text{GeV}_{8jets}^{\geq 3btags}$ & $60\text{GeV}_{9jets}^{\geq 3btags}$ & $60\text{GeV}_{10jets}^{\geq 3btags}$ & $80\text{GeV}_{8jets}^{\geq 3btags}$ & $80\text{GeV}_{9jets}^{\geq 3btags}$ & $80\text{GeV}_{10jets}^{\geq 3btags}$\\ \hline
   Exp & 4.8 \% & 2.6 \% & 1.0 \% & 2.9 \% & 1.2 \% & 0.4 \%  \\ \hline
   Sim & 5.7 \% & 2.8 \% & 1.2 \% & 3.0 \% & 1.3 \% & 0.37 \%  \\ \hline
   Sim2& 3.91\% &2.26 \% & 1.16\% & 2.63\% & 1.22\% & 0.475\% \\ \hline
 \end{tabular}
    \caption{\label{tab:bench7}} The final selection efficiency of all signal regions on our benchmark point. The numbers in the row of ``Exp" are given in the auxiliary Tab. II of Ref.~\cite{Aaboud_2017faq}. The results from our simulation are given in the row of ``Sim". We also give the corresponding values obtained by CheckMATE in the row of ``Sim2". 
  \end{table}

\section{ Search for paired dijet}  \label{app2}
Paper~\cite{Aaboud_2017nmi} is a RPV search for pair-produced resonances in four-jet final state at $\sqrt{s}=13$ TeV, with an integrated luminosity 36.7 fb$^{-1}$. 
There are two kinds of $UDD$ RPV vertices in this analysis, one is $t s d$ , the other is $tbs$. We will validate our analysis on the latter one and set the top squark fully decays to $bs$. Our signal process is $p p \to \tilde{t} \tilde{t}$ with a stop mass 500 GeV. Signal samples are generated using \madgraph interfaced to \pythia, with matching scale set to 100 GeV. \delphes is used to simulate the detector effects. The $b$-tag efficiency is chosen as 77\%, and the c-quark and light-quark mistaging efficiencies are 22.2\% and 0.77\%, respectively.

The upper limits for all signal regions are calculated by using Eq.~(\ref{uplimit1}).  
The results are shown in Table~\ref{tab:bench8}. 
\begin{table}[!h]\centering
 \begin{tabular}{|c|c|c|c|c|c|c|c|c|c|c|c|c|c} \hline
  $m_{\tilde{t}}$/GeV & 100 & 125 & 150 & 175 & 200 & 225 & 250 & 275 & 300 & 325\\ \hline
  $N_{limit}$ &199.38 & 462.98 & 704.55 & 524.46 & 774.34 & 353.98 & 443.62 & 357.81 & 216.86 & 202.32 \\ \hline
   $m_{\tilde{t}}$/GeV   & 350 & 375 & 400 & 425 & 450 & 475 & 500 & 525 & 550 & 575 \\ \hline
   $N_{limit}$   & 222.51 & 149.15 & 171.53 & 271.46 & 196.82 & 135.86 & 112.20 & 107.97 &98.94 & 100.30 \\ \hline
   $m_{\tilde{t}}$/GeV   & 600 & 625 & 650 & 675 & 700 & 725 & 750 & 775 & 800 & \\ \hline
   $N_{limit}$   & 86.13 & 90.34 & 59.30 & 43.49 & 46.46 & 54.89 & 48.76 & 70.50 & 78.74 & \\ \hline
  \end{tabular}
  \caption{\label{tab:bench8} New physics upper limits for all signal regions.}

\end{table}

According to the Table~I of Ref.~\cite{Aaboud_2017nmi}, we present the corresponding event numbers as well as cut efficiencies for the experimental analysis (Exp), our analysis (Sim) and CheckMATE analysis (Sim2)  in Table~\ref{tab:bench9}.

\begin{table}[!h]\centering
 \begin{tabular}{|c|c|c|c|c|c|} \hline
  & Total & Trigger & $\Delta {R_{min}}$ & Inclusive selection & $b$-tagged selection \\ \hline
  Exp & 18400 (100\%) & 11900 (64.67\%) & 2470 (13.42\%) & 253 (1.38 \%) & 65 (0.35\%)\\ \hline
  Sim & 19959 (100\%) & 13659 (68.44\%) & 2706 (13.56\%) & 211 (1.06 \%) & 80 (0.40\%)\\ \hline
  Sim2& 13190 (100\%) & 8429  (63.91\%) & 1764 (13.38\%) & 146 (1.11 \%) & 31 (0.24\%)\\ \hline
 \end{tabular}
 \caption{\label{tab:bench9} The analysis cut flows in the experimental paper (Exp), from our simulation (Sim) and from CheckMATE (Sim2).  Benchmark point with $m_{\tilde{t}}=500~{\rm GeV}$ is chosen.}
\end{table}

\section{Search for energetic muiltijet final state }  \label{app3}

In Ref.~\cite{ATLAS-CONF-2016-057}, the massive supersymmetric particles in multijet final states are searched. 
To validate our recast, we consider the gluino direct decay model as adopted in this experimental analysis. 

We generate $p p \to \tilde{g} \tilde{g}$ with \madgraph, then $\tilde{g}$ fully decays to $UDD$ quarks in \pythia, where the mass of $\tilde{g}$ is set to $1.1$ TeV. The upper limit can be calculated using Eq.~(\ref{uplimit1}) as before. 
The results are given in Table~\ref{tab:bench10}.  
The cuts flows of the analysis on the benchmark points have been provided in the Table~4 of Ref.~\cite{ATLAS-CONF-2016-057}. For comparison, we present both the experimental results (Exp) and our simulated results (Sim) in Table~\ref{tab:bench11}. It is not direct to implement the jet substructure analysis in CheckMATE. So we do not provide the corresponding results from CheckMATE in this recast.

\begin{table}[!h]\centering
 \begin{tabular}{|c|c|c|c|c|} \hline
 Signal Region & 4jSRb1 & 4jSR & 5jSRb1 & 5jSR \\ \hline
 Background & $61 \pm 10 $ & $151 \pm 15 $ & $18.2 \pm 4.2$ & $51.4 \pm 7.7 $ \\ \hline
 Observed & 46 & 122 & 30 & 64 \\ \hline
 Upper limit & 26.05 & 46.03 & 29.6 & 43.51 \\ \hline
 \end{tabular}
 \caption{\label{tab:bench10} The number of background events and their uncertainties as well as the observed event numbers in signal regions are provided in the Table 2 of Ref.~\cite{ATLAS-CONF-2016-057}. In the last row, we calculate the new physics upper limit for each signal region. }
\end{table}

\begin{table}[!h]\centering
 \begin{tabular}{|c|c|c|c|c|c|c|} \hline
   & Trigger & $p^{lead}_T>440\text{GeV}$ & $n_jet \geq 4$ & $M^{\Sigma}_J > 0.8$TeV & $|\Delta\eta_{12}| < 1.4$ & $b$-tag \\ \hline
   Exp & 2401 & 2236(93.13\%) & 1159(48.27\%) & 63.3(2.63\%) & 56.6(2.36\%) & 43.3(1.80\%) \\ \hline
   Sim & 2401 & 2107(87.76\%) & 1280(53.31\%) & 87.34(3.64\%) & 45.94(1.91\%) & 39.22(1.63\%) \\ \hline
 
 \end{tabular}
 \caption{\label{tab:bench11} Cut flow of the gluino direct decay model with $m_{\tilde{g}}=1100$ GeV in the experimental analysis (Exp) and from our simulation (Sim).   }
\end{table}

\end{appendix}

\phantomsection
\addcontentsline{toc}{section}{References}
\bibliographystyle{jhep}
\bibliography{refer}

\end{document}